# Experimental demonstration of diffusion limitations on resolution and SNR in MR microscopy

Benjamin M. Hardy[1,2], Yue Zhu[3], Kevin D. Harkins[2,5], Bibek Dhakal[1,2], Jonathan B Martin[2,5], Jingping Xie[2,4], Junzhong Xu[1,2,4,5], Mark D. Does[2,5], Adam W. Anderson[2,4,5], John C. Gore[1,2,4,5]

[1]Department of Physics and Astronomy, Vanderbilt University, Nashville, TN 37232, USA

[2] Vanderbilt University Institute of Imaging Science, Vanderbilt University Medical Center, Nashville, TN 37232, USA

[3]MR Engineering, GE Healthcare, Waukesha, WI 53188, USA

[4]Department of Radiology and Radiological Sciences, Vanderbilt University Medical Center, Nashville, TN 37232, USA

[5]Department of Biomedical Engineering, Vanderbilt University, Nashville, TN 37232, USA

\* **Corresponding author**: Address: Vanderbilt University, Institute of Imaging Science, 1161 21st Avenue South, AA 1112 MCN, Nashville, TN 37232-2310, United States.

E-mail: Benjamin.m.hardy@vanderbilt.edu (Benjamin M. Hardy)

**Running Title**: Diffusion limitations on resolution and SNR in MR microscopy

Abstract

Purpose: MR microscopy is in principle capable of producing images at cellular resolution (< 10 µm), but various factors limit the quality achieved in practice. A recognized limit on the signal to noise ratio and spatial resolution is the dephasing of transverse magnetization caused by diffusion of spins in strong gradients. Such effects may be reduced by using phase encoding instead of frequency encoding read-out gradients. However, experimental demonstration of the quantitative benefits of phase encoding are lacking, and the exact conditions in which it is preferred are not clearly established. We quantify the conditions where phase encoding outperforms a readout gradient with emphasis on the detrimental effects of diffusion on SNR and resolution.

Methods: A 15.2T Bruker MRI scanner, with 1 T/m gradients, and micro solenoid RF coils < 1 mm in diameter, were used to quantify diffusion effects on resolution and signal to noise ratio of frequency and phase encoded acquisitions. Frequency and phase encoding's spatial resolution and SNR per square root time efficiency were calculated and measured for images at the diffusion limited resolution. The point spread function was calculated and measured for phase and frequency encoding using additional constant time phase gradients with voxels 3-15 µm in dimension.

Results: The effect of diffusion during the readout gradient on SNR was experimentally demonstrated. The achieved resolutions of frequency and phase encoded acquisitions were measured via the point-spread-function and shown to be lower than the nominal resolution. SNR per square root time and actual resolution were calculated for a wide range of maximum gradient amplitudes, diffusion coefficients, and relaxation properties. The results provide a practical guide on how to choose between phase encoding and a conventional readout. Images of excised rat spinal cord at 10 µm x 10 µm in-plane demonstrate phase encoding's benefits in the form of higher measured resolution and higher SNR than the same image acquired with a conventional readout.

Conclusion: We provide guidelines to determine the extent to which phase encoding outperforms frequency encoding in SNR and resolution given a wide range of voxel sizes, sample, and hardware properties.

# 1. Introduction

The prospect of obtaining structural information at microscopic scale was mentioned in Lauterbur's revelation of the principle of MRI when he wrote "Zeugmatographic techniques should find many useful applications in studies of the internal structures, states, and compositions of microscopic objects" [1]. Mansfield and Grannell also stated that microscopic imaging of biophysical samples with magnetic resonance was possible [2]. Early studies by Hedges [3] identified the potential value and some challenges of MR microscopy. Ubiquitous in human and preclinical imaging, MRI provides deep 3D structural and functional tissue information with ≈ mm resolution. MR acquisitions with voxel dimensions less than ≈ 100 µm are typically referred to as magnetic resonance microscopy (MRM). MRM offers the many forms of contrast intrinsic to MR and may resolve cells without compromising cell viability [4–7]. In comparison, the poor penetration of light limits optical fluorescence microscopy to surface-level imaging and histology destroys the integrity of samples after processing. Thus, MRM is a prime candidate for in vivo, deep tissue imaging of intact cell systems.

With increased field strengths, strong gradients, cryogenically cooled RF coil arrays, and intelligent acquisition, reconstruction and post-processing, the practical limits of attainable resolution of MR acquisitions will continue to be challenged. Hedges acquired images on aquatic specimens with voxel volumes ≈ 0.15 nL with 35 µm in plane resolution [3], while Aguayo et al. imaged toad ova with 16 x 27 µm in-plane resolution in a 4 min acquisition at 9.5T [8]. Since then, the highest isotropic resolution claimed with MRM is 2.7 µm with constant-time 3D phase encoded gradients, 9.4T field, 100 µm diameter microcoil, and a frozen 1:1 water glycerin paramagnetically doped sample in 52 hours [9]. Other notably high-resolution images include 3.7 x 3.3 x 3.3 µm resolution images of polymer beads at 9T, acquired with 3D phase encoded gradient amplitudes up to 6 T/m in ≈ 30 hours [10].

Although phase encoding (PE) is claimed to be a superior acquisition scheme for MRM [11–16], not all acquisitions beyond 10 µm in the literature have used this approach. Most images of biological samples have been acquired with either Fast Low Angle Shot (FLASH) or Rapid Acquisition with Refocused Echoes (RARE) sequences with conventional frequency encoding (FE) in 1 dimension. For example, mammalian neurons were imaged with a 200 µm surface coil, 3 T/m gradients, 14.1T, at 4.7 µm isotropic resolution with a FLASH sequence in ≈ 22 hours [6]. Single mammalian myofibers at 6 µm isotropic, were imaged with FLASH and spin echo (SE) acquisitions with up to 25 T/m gradients, at 14.1T field strength, with a 200 µm surface coil in ≈ 30 hours [5]. More recently, plant root nodules were resolved with a 1.5 mm solenoid at 7 μm isotropic, 22.3T, and up to 2.88 T/m gradients in ≈ 30 hours [17].

The intrinsic signal and resolution limits of MR have been extensively discussed and modeled previously [3,16,18–22]. One of the primary limiting factors of MRM signal and resolution is diffusion. Callaghan and Eccles were two of the earliest to quantify the limits of MRM signal, resolution, and diffusion effects [23–30]. Diffusion during the readout gradient was shown to broaden the linewidth and attenuate signal. McFarland quantified sampling, diffusion, and $T_2$ contributions to the measured resolution of phase and frequency encoding via the point-spread-function (PSF) [12].

The discussion surrounding diffusive broadening led to the proposition of using PE in each dimension for MRM. Gravina et al [14] and Choi et al [13] demonstrated theoretically the advantage of constant time imaging's (3D PE but with 1 sample per repetition) use for imaging solids and its advantages over conventional readout acquisitions in signal and resolution. Elaborating on PE's advantages, Webb simulated optimal phase encoding steps for varying gradient strengths and sample diffusion properties to provide recommendations for improved PE resolution [11].

Although Gravina and Cory discuss achievable resolution given diffusive broadening, they ignore diffusive attenuation in their SNR comparison. Choi and coworkers compare the calculated PSF of phase and readout encoded experiments. They mention there is a threshold where PE begins to outperform frequency encoding in terms of time requirements for constant SNR and resolution. However, they do not clarify the experimental conditions in which PE has narrower resolution or is more SNR efficient.

Thus, the goal of this report is to demonstrate the conditions in which diffusion attenuates signal and degrades resolution in a conventional readout. We aim to guidelines to determine when PE outperforms readout-based acquisitions like FLASH or RARE. FLASH is chosen as the exemplar readout-based frequency encoded (FE) acquisition for comparison, although the results may be generalized for other readout acquisitions to some extent. The SNR per square unit time and resolution of PE are compared to that of FE, given a wide range of maximum gradient strengths, relaxation, and diffusion properties of the sample.

We measured the PSF for phase and frequency encoded acquisitions to confirm analytical models of broadening due to relaxation and diffusion. We experimentally demonstrate diffusive attenuation over shrinking voxel dimensions. The SNR per square unit time of FE and PE sequences are compared, including the effects of diffusive attenuation with varying gradient strengths, T$_2$, and diffusion values. Finally, the benefits of PE in improved SNR and resolution are demonstrated with 10 μm in-plane images of excised rat spinal cord acquired at 15.2T with max 1 T/m gradients.

## 2. Theory

We aim to compare the imaging efficiency (SNR per square unit time) in a FE versus a PE sequence assuming realistic sequence and sample parameters.

2.1 Imaging Efficiency excluding Diffusion and Relaxation

The SNR per voxel a 3D acquisition may be written as

$$SNR/voxel \propto \Delta x\, \Delta y\, \Delta z\, \sqrt{N_{acc} N_x N_y N_z t_{acq}}\,, \tag{1}$$

where Δx, Δy, Δz are the voxel sizes in each dimension, N$_{acc}$ is the number of accumulated signals averaged, N$_x$, N$_y$, N$_z$ are the number of samples in each dimension, and t$_{acq}$ is the signal sampling duration. For a FE experiment, t$_{acq}$ is the inverse of the receiver bandwidth, t$_{acq}$ = 1/BW and equation 1 may be further simplified by including the duration of the readout gradient, T$_{read}$ = N$_x$/BW = N$_x$t$_{acq}$ (fig. 1 left). The total time for a frequency encoded experiment is T$_T$ = T$_R$ N$_y$ N$_z$ N$_{acc}$ as every repetition time (T$_R$) includes N$_x$ samples. For a PE sequence, t$_{acq}$ = N$_t$/BW as after each excitation N$_t$ samples of the FID may be sampled within the T$_R$ (fig.1 right). Total time for the PE sequence is T$_T$ = T$_R$ N$_x$ N$_y$ N$_z$ N$_{acc}$ as only 1 N$_x$ sample is acquired per T$_R$. The SNR does not continue to increase with t$_{acq}$ beyond a limiting value as the signal itself decays exponentially with a time constant of T$_2^*$. A more complete comparison includes relaxation effects (see 2.2).

The interplay between SNR, time, and resolution is already apparent in equation 1. With fixed field of view (FOV) and bandwidth, doubling the isotropic resolution of a 3D scan shrinks the voxel volume by a factor of 8 and the SNR drops by a factor of 8$^{-1/2}$ as the number of samples is also doubled. Therefore, to maintain SNR with doubled resolution, the experiment must be averaged 8-fold, taking 32 (8-fold * 2 y samples * 2 z samples) or 64 times as long for frequency or phase encoding respectively. Considering the time differences in these acquisitions, it is useful to compare the SNR per square unit of time or the imaging efficiency defined as

$$\eta = \frac{SNR}{\sqrt{T_T/T_1}}, \tag{2}$$

where $T_T$ is the total time of the acquisition normalized by the spin-lattice relaxation of the sample $T_1$. Using the proportion in eq. 1, setting $N_x t_{acq} = T_{read}$, and plugging in the $T_T$ for a frequency and phase encoded acquisition, the ratio of imaging efficiency of FE to PE is

$$\eta^{FE}/\eta^{PE} = R^{FE/PE} \propto \sqrt{T_{read}/t_{acq}}. \tag{3}$$

Thus, when $T_{read} = t_{acq}$ the two acquisitions are equally efficient. For porous structures and solids, constant time or single point imaging [15,31–34] is used because $t_{acq}$ is limited by very short $T_2^*$ values and $N_t$ is limited to a few points. A more complete comparison of the acquisitions is required for resolution of biological samples at microscopic resolution by including signal loss to $T_2^*$, signal attenuation due to diffusion, and their respective resolution degradations.

## 2.2 Imaging Efficiency including $T_2^*$ and Steady State

To include the effects of sequence timing and relaxation of the sample on SNR we continue from equation 18 in Link et al [35]. The FE imaging efficiency depends on the echo time (TE), $T_R$, and sample relaxation properties as

$$\eta^{FE} = SNR_0 e^{-TE/T_2^*} \frac{1 - e^{-T_{read}/T_2^*}}{\sqrt{T_{read}/T_2^*}} \frac{(1-E_1)\sin(\arccos E_1)}{(1-E_1^2)\sqrt{T_R/T_1}}, \tag{4}$$

$$where\ E_1 \equiv e^{-T_R/T_1}.$$

$SNR_0$ in equation 4 is scanner and sample dependent and is defined as $S_0/N_0$ where $S_0$ is the equilibrium magnetization measured immediately after a 90-degree pulse and $N_0$ a reference noise metric dependent on the bandwidth, hardware, and sample properties. $T_2^*$ is the decay time of the signal due to $T_2$ and field inhomogeneities. Equation 4 includes the optimal Ernst angle for steady state acquisitions. For PE, $\eta^{PE}$ is the same as eq. 4, only $T_{read}$ is replaced with $t_{acq}$ [36]. The terms in eq. 4 that include $T_2^*$ have a maximum when $t_{acq}$ or $T_{read} = 1.25 T_2^*$ [37].

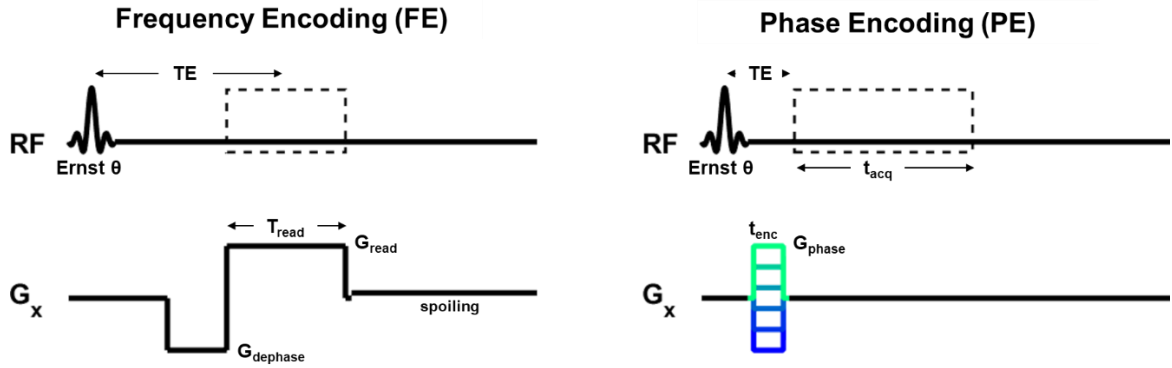

Figure 1: Pulse sequence diagrams indicating the key differences in the two acquisitions compared. A conventional readout gradient frequency encodes the first dimension. The phase encoded acquisition acquires data while the gradients are off after the phase encoding period, $t_{enc}$. Both sequences' RF pulses use the Ernst angle.

## 2.3 Diffusion's Effects on Signal

Diffusion in non-uniform magnetic fields causes spin dephasing which attenuates signal [38,39]. During the frequency encoding readout gradient, the attenuation is

$$A_{FE}(D, G_{read}, T_{read}) = e^{-\frac{1}{12}DG_{read}^2 T_{read}^3}, \tag{5}$$

where D is the self-diffusion coefficient of spins in the sample (in the read direction), and $G_{read}$ the strength of the readout gradient. Equation 5 assumes that $G_{dephase} = G_{read}$. The factor of 1/12 will change depending on the precise shape in time of the dephase and readout gradients (Appendix A). Similarly, the attenuation during phase encoding may be written as,

$$A_{PE}(D, G, t_{enc}) = e^{-\frac{1}{3}\gamma^2 DG_{phase}^2 t_{enc}^3}. \tag{6}$$

The PE gradient waveform polarity is fixed for every repetition thus waveform is fixed, and the exponent is varies slightly from FE. It is clear when $T_{read} \gg t_{enc}$, the attenuation for phase encoding may be much less for PE.

## 2.4 Diffusion's Effects on Imaging Efficiency

By combining equations 2 and 4-6, the ratio of FE to PE's imaging efficiency is altered by the influence of diffusion on signal but may be defined as,

$$R^{FE/PE} = \frac{\eta^{FE}}{\eta^{PE}} \frac{A_{FE}}{A_{PE}}. \tag{7}$$

Substituting equation 4 for FE and PE into equation 7 and cancelling like-terms,

$$R^{FE/PE} = \sqrt{\frac{t_{acq}}{T_{read}} \frac{e^{TE^{FE}/T_2^*}}{e^{TE^{PE}/T_2^*}} \frac{[1 - e^{-T_{read}/T_2^*}]}{[1 - e^{-t_{acq}/T_2^*}]} \frac{A_{FE}}{A_{PE}}}. \qquad (8)$$

Figure 2 plots curves using equation 8 with varying gradient strengths and nominal resolution. The following sample and acquisition parameters are fixed in the plot: $T_2^* = 10$ ms, FOV = 2 mm, $G_{read} = G_{phase} = 0.1 - 10$ T/m, BW = $\gamma$FOV$G_{read}/2\pi$ where $\gamma$ is the gyromagnetic ratio, $t_{acq} = T_{read}$, where $T_{read} = N_x$ BW$^{-1}$, and $N_x$ is determined by the FOV and resolution, and $TE^{FE} = TE^{PE}$. At larger voxel sizes, R = 1. In smaller voxel sizes, attenuation of signal by diffusion in the readout gradient dominates and the advantages of PE become increasingly apparent in more time efficient acquisitions. The effects of longer read durations may be ameliorated by using higher gradient strengths to reduce signal losses in smaller voxels.

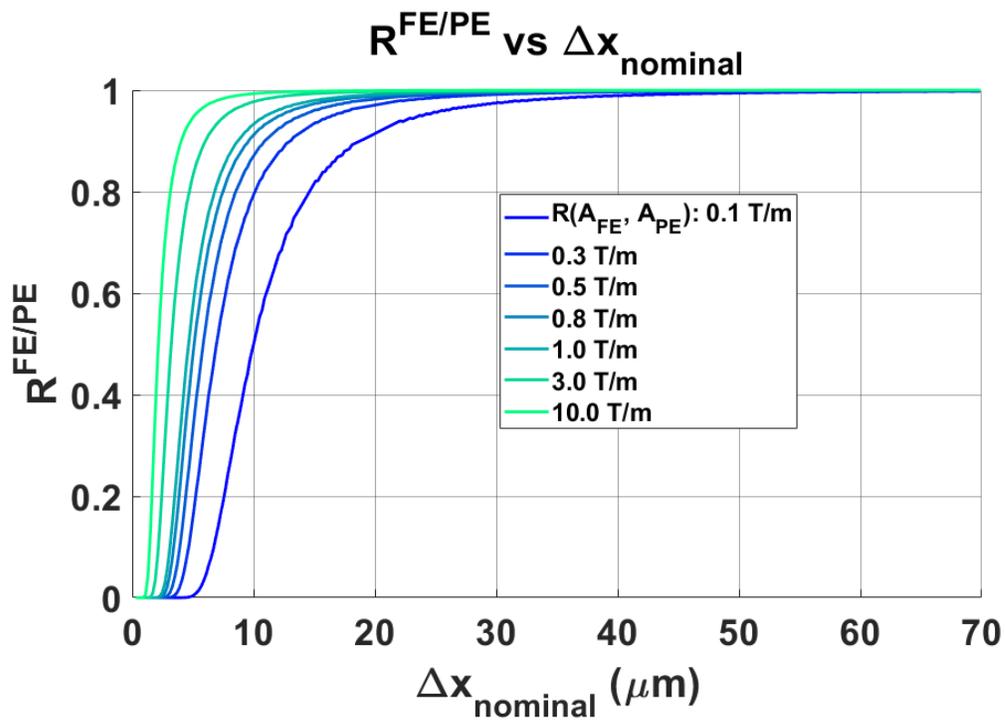

Figure 2: $R^{FE/PE}$ in a simulated sample and imaging scheme where $t_{acq} = T_{read}$, $T_2^* = 12$ ms, and diffusion is that of free water. For larger voxel sizes, R converges on 1. In smaller voxels, the readout gradient rapidly attenuates signal, and PE outperforms FE with more time efficient acquisitions.

## 2.5 Diffusion's Effects on Resolution

Another important consideration for high-resolution MRM is the effect of diffusion on the spatial resolution actually achieved. As has been previously reported, PE suffers less line broadening and thus offers an intrinsically narrower PSF [11,12,15,32,34,40]. To compare the differences between the PSF for phase and frequency encodes we use a numerical calculation and measurements of the PSF using an MR sequence. For clarity, nominal resolution refers to the resolution determined by the FOV and number of samples, such that $\Delta x_{nominal}$ = FOV/$N_x$. $\Delta x_{actual}$ refers to the resolution predicted by our transfer function analysis outlined below in 2.5.1. Finally, $\Delta x_{measured}$ is the resolution returned by the measurement method outlined in 2.5.2

### 2.5.1 Analytical Point Spread Function Calculation

We follow closely McFarland and Webb's approach [11,12]. The actual resolution for frequency encoding may be approximated as the FWHM of the PSF or

$$\Delta x_{actual}^{FE} = FWHM[FT^{-1}(MTF_{sampling} MTF_{relaxation} MTF_{diffusion}^{FE})] \quad (9)$$
$$= FWHM[PSF_{sampling} * PSF_{relaxation} * PSF_{diffusion}^{FE}],$$
$$where * denotes\ convolution.$$

MTF is the modulation transfer function and $FT^{-1}$ is the inverse Fourier transform. It should be noted that $PSF_{sampling}$ includes finite sampling and truncation of the MR signal resulting in the well-known sinc function [41]. For phase encoding, $T_2^*$ decay does not contribute to the resolution degradation because the echo time is kept constant for all values of the phase encoding gradients. Thus, the actual resolution of phase encoding may be approximated as

$$\Delta x_{actual}^{PE} = FWHM[FT^{-1}(MTF_{sampling} MTF_{diffusion}^{PE})] \quad (10)$$
$$= FWHM[PSF_{sampling} * PSF_{diffusion}^{PE}].$$

For both phase and frequency encoding $PSF_{sampling}$ is

$$PSF_{sampling} = \Delta k \frac{sin(\pi N_x \Delta kx)}{sin(\pi \Delta kx)}, where\ \Delta k = 1/FOV. \quad (11)$$

Considering relaxation during frequency encoding,

$$PSF_{relaxation} = FT^{-1}[e^{-t/T_2^*}] = FT^{-1}[e^{-k/\gamma G_{read} T_2^*}] = \quad (12)$$

$$\frac{a}{\pi(a^2+x^2)}, where\ a = 1/\gamma G_{read} T_2^*.$$

McFarland shows that the diffusion contributions to the PSF in frequency encoding may be written as

$$PSF_{diffusion}^{FE} = FT^{-1}[MTF_{diffusion}^{FE}] = FT^{-1}[e^{-k^3 D/3\gamma G_{read}}]. \quad (13)$$

Webb derived the PSF contributions from diffusion for phase encoding with variable gradients,

$$PSF_{diffusion}^{PE} = e^{-3x^2/4Dt_{enc}}. \quad (14)$$

In figure 3, the actual resolution is numerically calculated for phase and frequency encoding by using equations 9-14.

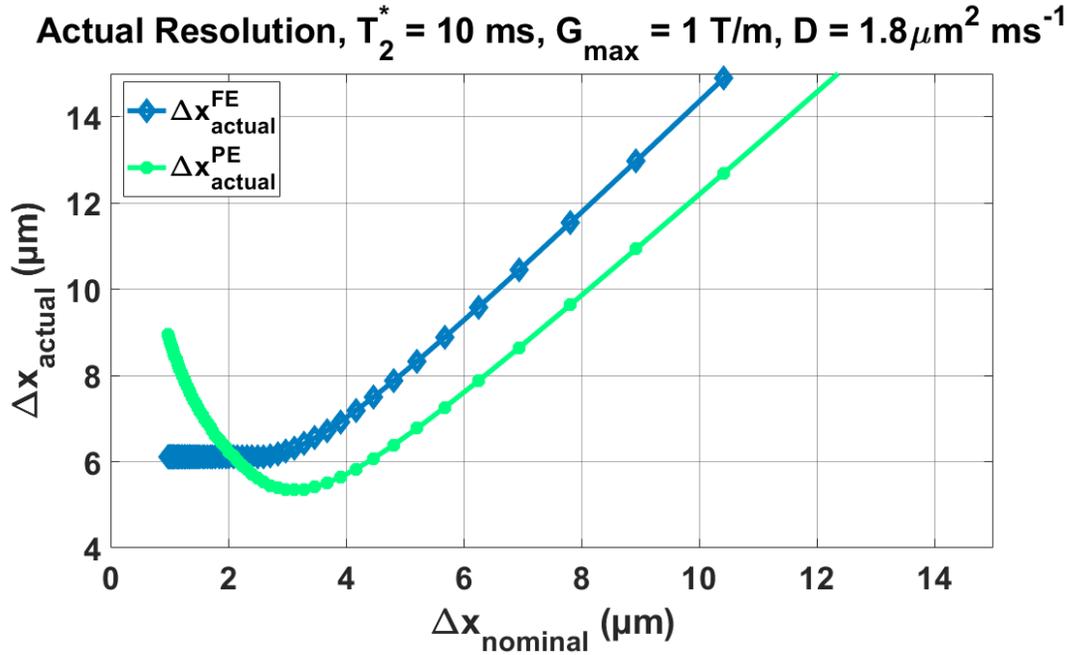

Figure 3: Equations 9 and 10 demonstrate the clear resolution benefit of phase encoding. Relaxation does not contribute to PE's PSF and diffusion's effects are lessened, phase encoding may achieve a higher achievable resolution. With voxels smaller than 3 µm, and a fixed gradient, phase encoding's actual resolution begins to degrade due to $t_{enc}$ increasing.

## 2.5.2 Measurement of Point Spread Function with Imaging

To experimentally measure the PSF, we use a constant time gradient sequence. Signals are acquired using extra phase encoding gradients in the respective dimension to be measured, followed by an inverse Fourier transform. This method was previously developed and demonstrated for EPI-based acquisitions by Robson et al and Zeng and Constable [42,43].

A full k-space trajectory is traversed for each value k' of the additional phase encoding gradients. The k-space data are inverse Fourier transformed along each spatial dimension forming an image for each k'. This is followed by an inverse Fourier transform along the k' direction at each pixel. The resulting profile is the spin-density weighted point spread function. An example is provided in figure 4.

The measured resolution may then be defined as

$$\Delta x_{measured} = FWHM[FT^{-1}(S(x,y,k'))], \quad (15)$$

where x and y denote the pixel spatial location, and *S* denotes sample image from each k' encoding. To measure the PSF precisely, k' space must be adequately sampled. Adequate sampling may be achieved by increasing the number of k' encoding points ($N_{PSF}$) or increasing Δk' by increasing the gradient (Δk' = 1/$FOV_{PSF}$). A compromise between imaging time and measurement accuracy was determined by setting $FOV_{PSF}$ = ¾ FOV and $N_{PSF}$ = 2$N_x$. see supplementary figure 3 for more information.

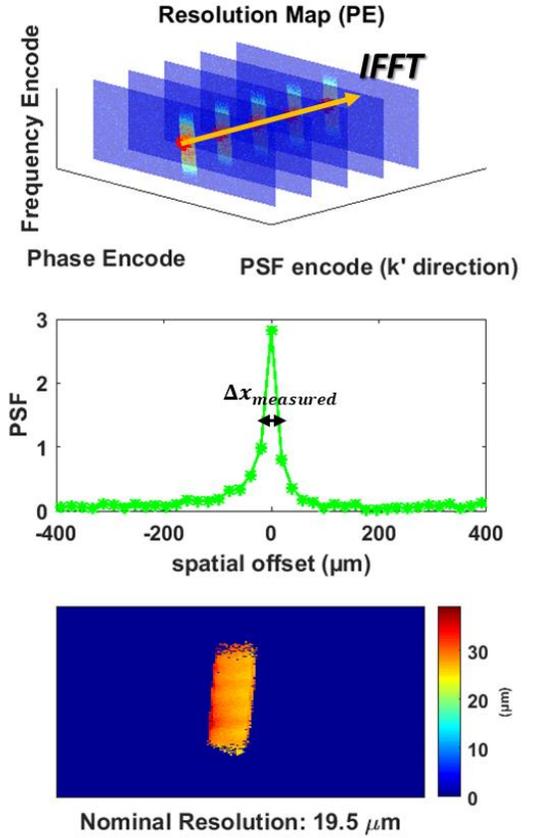

Figure 4: A resolution map is obtained by plotting the FWHM of the PSF for each pixel within the sample. The inverse Fourier transform is performed for each pixel along the k' direction to render the PSF. Then the FWHM of the PSF for each pixel is plotted back onto the masked sample. In this case, a $CuSO_4$ filled capillary.

# 3. Methods

### 3.1 MRI Equipment, Hardware, and phantom

All data were acquired with a Bruker (Billerica, MA) 15.2T Biospec imaging spectrometer equipped with max 1 T/m magnetic field gradients. The Avance III console used ParaVision 6.0.1. The bore is 6 cm in diameter within the gradients with $3^{rd}$ order $B_0$ shims. For SNR and PSF measurements a small solenoid was used. The solenoid was a copper, 6 turn, 1.5 mm diameter coil with a wire diameter of 300 µm coated in polyurethane. The tune and match circuit consisted of the solenoid, 2 variable capacitors, and 2 fixed capacitors (see supplementary figure 2). A Bruker 20 mm cryoprobe was used to image excised tissue described in 3.3. The phantom for resolution and SNR experiments was a 6 mM $CuSO_4$ solution in glass capillaries of varying sizes (Drummond Scientific Company, Broomall, PA). The capillaries were filled via the capillary effect and sealed using a wax pen.

## 3.2 SNR measurements

SNR was measured by selecting 2 regions of interest across magnitude images of the experimental sample. The mean signal region consisted of pixels within the sample and the noise region consisted of areas at the corners of the image with no sample present. The magnitude image values were plotted on a log-scale to ensure noise and sample regions were visually free from distortion or artifacts. Magnitude images were used so a correction was applied following the work of Henkelman [44,45] in order to ensure low signal images did not overestimate SNR.

## 3.3 Experiments

To demonstrate the effects of diffusion at high resolution, several experiments and simulations were performed.

### 3.3.1 Demonstration of Diffusion in a Read-Out Gradient

To demonstrate attenuation due to diffusion during the readout gradient, images at varying resolution in the frequency encoding dimension were acquired. The resolution was varied by changing the number of readout points which lengthened $T_{read}$. TE was fixed for every image by using the minimum TE allowed for the highest resolution image. The following parameters were fixed for each nominal resolution: the bandwidth at 40 kHz, $G_{read}$ = 0.47 T/m, $N_{acc}$ = 9, FOV = 2 x 2 mm, TE/TR = 14.7/80 ms, Ernst angle of 47.9°, and slice thickness, $\Delta z$ = 2 mm. Each scan took ≈ 23 seconds. The sample was a 0.4 mm inner diameter 0.9 mm outer diameter capillary tube filled with 6 mM $CuSO_4$ solution.

The SNR was fit to the following equation formulated from equation 1,4 and 7,

$$SNR_{2D}^{FE} = SNR_0 \, \Delta x \Delta y \Delta z \sqrt{N_{acc} N_y} \, e^{-TE/T_2^*} \frac{1 - e^{-T_{read}/T_2^*}}{\sqrt{T_{read}/T_2^*}} \frac{(1 - E_1) \sin(\arccos E_1)}{(1 - E_1^2)} A_{FE} \quad (16)$$

$$\text{where } A_{FE} = e^{[-\varepsilon \gamma D G_{read}^2 T_{read}^3]}.$$

In equation 16, TE is fixed, $T_{read} = T_s \, FOV_x/\Delta x$, and ε depends on the duration and amplitude of the dephase gradient (Appendix A), in this case $G_{read} \neq G_{dephase}$ so $\varepsilon \neq 1/12$. The data were fitted allowing $SNR_0$ and ε to vary. Given $G_{read}$, the bandwidth was calculated for FOV = 2 mm, using BW = $FOV \gamma G_{read}/2\pi$.

### 3.3.2 Demonstration of Imaging Efficiency Loss Due to Diffusion

To demonstrate the steep, drop off in the ratio of imaging efficiencies due to diffusion, a simple experiment was executed. The resolution was varied for the FE and PE from 15 to 10 μm. The resolution was varied by changing the number of readout points which lengthened $T_{read}$. The following parameters were fixed for each nominal resolution: the bandwidth at 40 kHz, $G_{read}$ = 0.47 T/m, $N_{acc}$ = 1, FOV = 2 x 2 mm, TE/TR = 7/20 ms, and slice thickness, $\Delta z$ = 0.6 mm. Each FE acquisition was ≈ 2.6 s and the PPE scans ran from 5 to 8 min. The sample was a 0.4 mm inner diameter 0.9 mm outer diameter capillary tube filled with 6 mM $CuSO_4$ solution. The microcoil was a 6 turn 1.5 mm diameter coil with a wire diameter of 300 μm coated in polyurethane. $G_{phase}$ was set to $G_{read}$ by varying $t_{enc}$ for each resolution. $G_{dephase}$ was set to $G_{read}$ by varying the dephasing duration for each resolution.

### 3.3.3 Measuring Resolution broadening

To demonstrate broadening contributions of diffusion and $T_2^*$ to the actual resolution of the experiment, the PSF was acquired at varying resolutions with the microcoil using theory outlined in 2.5.2. To measure the actual resolution in the frequency encoded dimension, the number of readout points was incrementally increased from 67, 100, and 125 points with a FOV of 1 x 1 mm and $N_y$ set to 32. There were 134, 200, and 250 $N_{PSF}$ encoding points. For the phase dimension, the same FOV and $N_{PSF}$ points were used only the PSF was acquired in the y direction and $N_x$ was 32. For each acquisition, the following parameters were fixed: $G_{read}$ = 0.56 T/m, bandwidth = 24 kHz, $N_{acc}$ = 9, TE/TR = 16.4/80 ms, and a slice thickness of 1 mm.

### 3.3.5 Excised Rat Spinal Cord Images

To demonstrate a scenario in which PE is clearly advantageous in SNR and resolution, images of an excised rat spinal cord were acquired. The spinal cord was divided near the lower lumbar/upper sacral region. The tissue was prepared for imaging by first fixing in a 10% Phosphate buffer solution (PBS). It was then soaked overnight in 2 mM Gadolinium (Gd) and 0.1 mM sodium azide ($NaN_3$). To remove excess signal, the tissue was placed in Fomblin (Solvay Brussels, Belgium) and a microcentrifuge tube for imaging.

The following parameters were fixed for the two acquisitions: FOV = 5 x 5 mm, 512 x 512 acquisition matrix, resulting in 9.8 x 9.8 µm resolution, 250 µm slice thickness. For the FE image a FLASH sequence was used with TE/TR = 8.2/200 ms, an Ernst angle of 83.8°, BW = 50 kHz, $G_{read}$ = 0.23 T/m, $T_{read}$ of 10.2ms and $N_{acc}$ = 64. For the PE sequence, TE/TR = 2.5/30 ms, an Ernst angle of 83.8°, the BW of the receiver was 3.2 kHz, $N_s$ = 64, with $t_{acq}$ = 20 ms, and $t_{enc}$ = 1.75 ms. $G_{phase}$ was 0.69 T/m.

### 3.3.6 Microscopy recommendations Considering $T_2^*$, $G_{max}$, and $D$

To provide recommendations for imaging in the diffusive regime at very high resolution, equations 8, 9, and 10 were used to calculate $R^{FE/PE}$, $\Delta x_{actual}^{FE}/\Delta x_{nominal}$, and $\Delta x_{actual}^{PE}/\Delta x_{nominal}$ over varying diffusion, $T_2^*$ and $G_{max}$ values. The $T_{read}$ was determined by 3 fixed $G_{max}$ values 0.5, 1, and 2 T/m and a fixed FOV of 2 mm. $T_2^*$ values were 1, 10, and 30 ms. Diffusion values were 0.5, 1, and 2 µm² ms⁻¹. It is assumed that TR and TE are fixed for the two acquisitions. To simulate the most realistic case, $t_{acq}$ was set to the SNR ideal value of $1.25T_2^*$ [35,37]. The $\Delta x_{nominal}$ was calculated for two conditions: 1) When the PE is twice as time efficient as FE and 2) when the encoding returns actual resolutions 1.5x the nominal.

## 4. Results

### 4.1 Diffusion During the Read-Out

The experimentally measured SNR is plotted in figure 5 as red dots. By simply shrinking the voxel size, the SNR is expected to go down since there is less signal in a smaller voxel (red diamonds). The predicted SNR due to smaller voxels does not adequately characterize the measured loss in SNR as the resolution is increased experimentally. However, once the attenuative term (eq. 5) is added to the model, the fit (blue diamonds) follows the measured data (red dots).

With the diffusion attenuation term, $R^2 = 0.99$ and without it, $R^2 = 0.73$. $SNR_0$ best fit was 4.4e13 and $\varepsilon = 0.038$. The diffusion coefficient of the sample was measured to be 1.8 µm$^2$ ms$^{-1}$ based on a diffusion weighted image prior to the experiment. The $T_1$ and $T_2$ of the sample was ≈ 200 and 12 ms respectively.

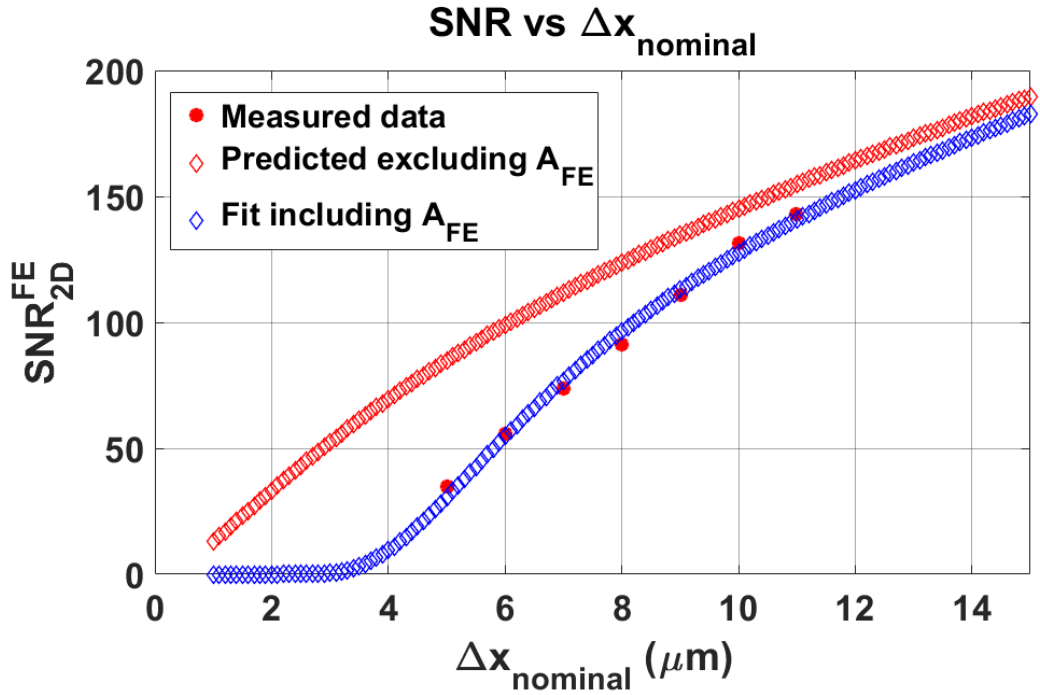

Figure 5: SNR of a FE acquisition was measured with increasing resolution in the readout dimension. Resolution was increased by fixing the FOV and increasing the number of samples acquired during the readout extending the duration of the readout gradient. The TE was fixed by using the minimum TE for the highest resolution scan. For comparison, the model was plotted with and without the diffusive attenuation term given the same initial SNR. At larger voxel sizes, the two models agree since the attenuation term is negligible for shorter read durations.

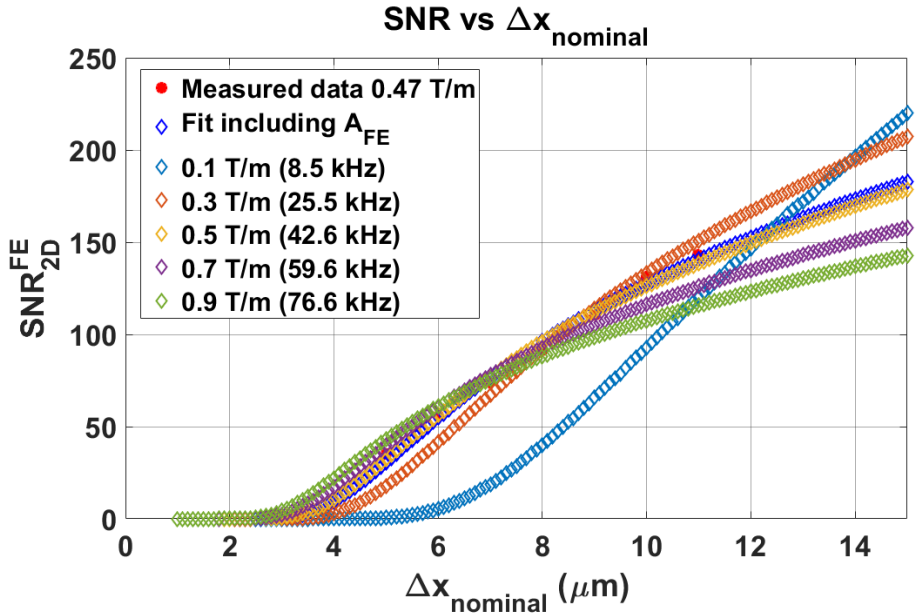

Figure 6: The simulated SNR for varying gradient strengths for very small voxels in a FE acquisition. For the smallest voxels, stronger gradients have higher SNR since the attenuation $\propto \exp(G^2 T^3)$ and longer readouts are penalized at high resolution. For larger voxels, the well know relationship of SNR $\propto BW^{-1/2}$ is apparent and weaker gradients produce higher SNR. Between the extremes care must be taken to choose optimal gradient strength.

## 4.2 Simulated $G_{read}$ and $T_{read}$ combinations

Using the fit in figure 5, it is useful to consider varying $T_{read}$ and $G_{read}$ combinations effects on SNR attenuation. For larger voxels, the well know relationship of SNR $\propto BW^{-1/2}$ holds, although a more precise description of this relationship is that SNR $\propto$ the time that the receiver is on [46]. When diffusion attenuates signal, stronger gradients have an SNR advantage. Given a chosen resolution in the readout dimension, there exists an optimal bandwidth and $G_{read}$ combination. To demonstrate this for the $CuSO_4$ phantom, $G_{read}$ and the bandwidth were varied with simulation while maintaining $SNR_0$ and $\varepsilon$ from the fitted data in figure 5. The results of varying $G_{read}$ are plotted in figure 6. Given $G_{read}$, the bandwidth was calculated for FOV = 2 mm, using BW = FOV$\gamma G_{read}/2\pi$. For voxels larger than 14 µm, the smallest BW has the largest SNR. For 13-9 µm, the 0.3 T/m gradient has the highest SNR. Between 9 and 6 µm, 0.5 T/m is optimal. Below 6 µm, the highest gradient has the highest SNR since $T_{read}$ is the shortest.

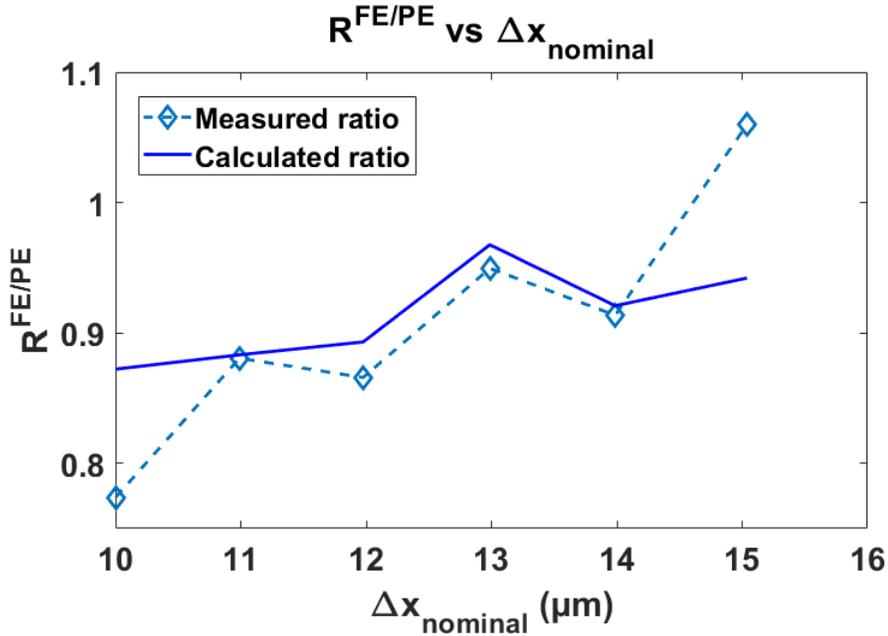

Figure 7: Experimental verification of equation 8. With a fixed readout gradient, diffusion begins to attenuate the SNR of a FE acquisition and PE becomes more efficient. For diffusion of free water, $T_2^* = 12$ ms, and $G_{read} = G_{phase} = 0.47$ T/m, R < 0.9 when $\Delta x_{nominal} = 11$ µm.

## 4.3 Demonstration of Imaging Efficiency Loss Due to Diffusion

From figure 2, it is expected that for gradients < 1 T/m $R^{FE/PE}$ will begin falling below 1 when $\Delta x_{nominal}$ is 10-20 µm. This is shown in figure 7 to experimentally verify equation 8. The experimentally measured ratio (diamonds) is plotted followed by the calculated ratio (solid line) given by equation 8. The acquisition durations varied from 5 to 3.4 ms. The calculated ratio is plotted using equation 8 and by plugging in the experimental parameters. For the PE sequence, the receiver bandwidth was fixed to 3.2 kHz for every acquisition. This resulted in small differences in $t_{acq}$ from $T_{read}$ since $t_{acq} = N_s/BW$ and depends on a finite number of samples, thus the calculated and experimental data do not vary smoothly.

## 4.4 Measuring Resolution Broadening

For the frequency encoding resolutions (blue histogram in fig. 8) the scan times were as follows: 51m, 1hr17m, 1hr36m, for 14.9, 10, and 8 μm respectively. The phase encoding resolutions (green histograms in fig. 8) had the following scan times: 1hr48m, 4hrs, 6hrs15m for 14.9, 10, and 8 μm respectively. The phase encoding experiment took much longer since changing $N_x$ and $N_{PSF}$ increased the time whereas for the frequency direction, only changing $N_{PSF}$ increased the time requirements.

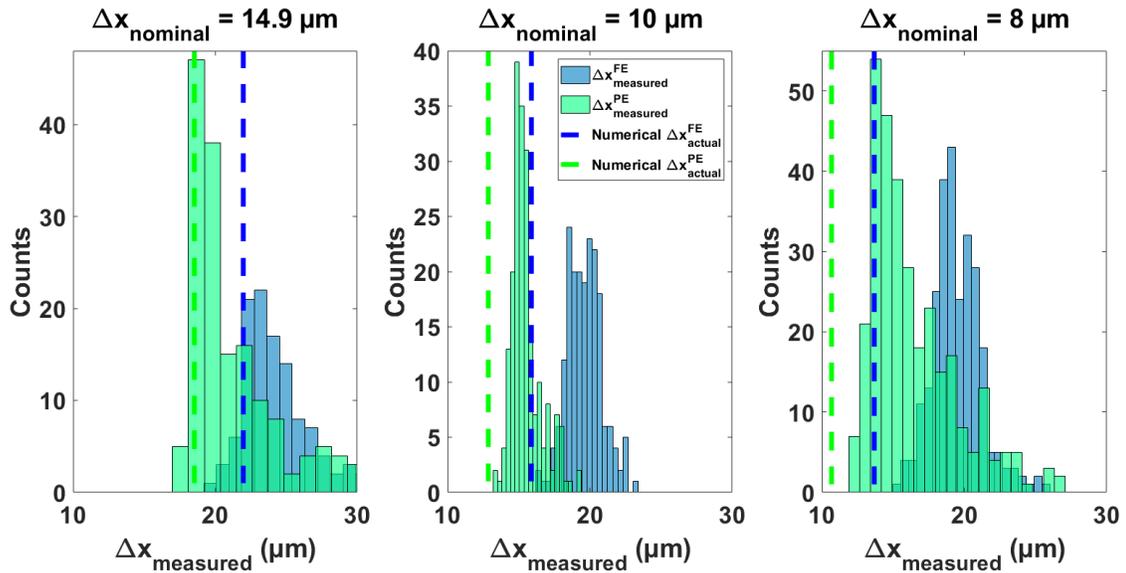

Figure 8: The numerical predictions of the actual resolution show agreement with the measured resolutions at larger voxel sizes 14.9 μm. However, as diffusion begins to degrade resolution, the additional phase encoding gradients from the measurement begin to contribute even more resolution broadening.

For $\Delta x_{nominal}$ = 14.9 μm, the mean of the $\Delta x_{measured}$ was 24.8 and 22.8 μm for FE and PE. The predicted $\Delta x_{actual}$ was 21.1 and 18.5 μm for FE and PE respectively. For $\Delta x_{nominal}$ = 10 μm, the mean of the $\Delta x_{measured}$ was 19.6 and 15.5 μm for FE and PE. The predicted $\Delta x_{actual}$ was 15.9 and 12.9 μm for FE and PE. For $\Delta x_{nominal}$ = 8 μm, the mean of the $\Delta x_{measured}$ was 19.5 and 16.4 μm for FE and PE. The predicted $\Delta x_{actual}$ was 13.7 and 10.7 μm for FE and PE.

The analytical predictions of the actual resolution (equations. 9 and 10) are included in figure 8 as the blue and green dashed lines. For each nominal resolution, phase encoding returns a better measured resolution. There is good agreement in the analytical measurement for the 14.9 μm voxel sizes. However, as the resolution is increased the measurement is worse. This may be due to the analytical model not accounting for the extra phase encoding gradients being used in the measurement which also contribute to resolution broadening.

## 4.5 Excised Rat Spinal Cord Images

For the FE FLASH image (figure 9, left side), the total imaging time was 1hr49m and SNR = 0.5. For the PE image (figure 9, right side), the total imaging time was 2hrs11m and SNR = 8.5. Each repeated scan for the FLASH sequence had SNR of roughly 0.5. Assuming ideal SNR averaging, the total SNR would be roughly 4 for 64 $N_{acc}$. From equation 8, the expected $R^{FE/PE}$ = 0.4 with the above acquisition parameters compared to ideal averaged SNR, $R^{FE/PE}$ is 0.47. The $T_2^*$ across the whole sample was ≈ 20 ms. The $T_1$ was measured on average ≈ 90 ms. The images are normalized to the max signal and masked.

## 4.6 Microscopy Recommendations Considering $T_2^*$, $G_{max}$, and $D$

To provide recommendation for imaging with optimal actual resolution and time efficiency, equations 8, 9, and 10 were used to determine the nominal resolutions at which PE is a clear choice for an imaging experiment. The results are presented in Table 1. An example use of this table is as follows. If a tissue sample of interest has a diffusion value of 1 $\mu m^2$ $ms^{-1}$ and $T_2^*$ of 10 ms and the MR scanner in use has a maximum gradient of 2 T/m. PE will be twice as time efficient as a FE acquisition at $\Delta x_{nominal}$ = 10 μm resolution. FE images with resolution higher than 10 μm will only suffer more time requirements and resolution broadening. Similarly, with D = 2 $\mu m^2$ $ms^{-1}$, $T_2^*$ = 10 ms, and $G_{max}$ = 0.5 T/m, frequency encoding will return $\Delta x_{nominal}$ = 10 μm images with 150% broadening, preventing resolution of details < 15 μm. With no increased time requirements, PE will only blur by 120%, resulting in images with 3 μm finer resolution.

| D (µm² ms⁻¹) | $T_2^*$ (ms) | G (T/m) | $\Delta x_{FE}/\Delta x_{nominal} = 1.5$ $\Delta x_{nominal}$ (µm) | $\Delta x_{PE}/\Delta x_{nominal} = 1.5$ $\Delta x_{nominal}$ (µm) | $R^{FE/PE} = 0.5$ $\Delta x_{nominal}$ (µm) |
|---|---|---|---|---|---|
| 0.5 | 1 | 0.5 | 33.5 | 3.0 | 6.4 |
|  |  | 1 | 18.7 | 2.4 | 4.1 |
|  |  | 2 | 10.5 | 1.9 | 2.9 |
| 0.5 | 10 | 0.5 | 7.5 | 3.0 | 3.8 |
|  |  | 1 | 5.1 | 2.4 | 3.1 |
|  |  | 2 | 3.6 | 1.9 | 10.1 |
| 0.5 | 30 | 0.5 | 5.6 | 3.0 | 13.2 |
|  |  | 1 | 4.2 | 2.4 | 5.1 |
|  |  | 2 | 3.2 | 1.9 | n/a |
| 1 | 1 | 0.5 | 35.2 | 3.8 | 7.2 |
|  |  | 1 | 19.8 | 3.0 | 4.8 |
|  |  | 2 | 11.4 | 2.4 | 3.4 |
| 1 | 10 | 0.5 | 8.8 | 3.8 | 4.8 |
|  |  | 1 | 6.1 | 3.0 | 4.0 |
|  |  | 2 | 4.4 | 2.4 | 9.8 |
| 1 | 30 | 0.5 | 6.8 | 3.8 | 12.6 |
|  |  | 1 | 5.2 | 3.0 | n/a |
|  |  | 2 | 4.0 | 2.4 | n/a |
| 2 | 1 | 0.5 | 36.0 | 4.8 | 8.3 |
|  |  | 1 | 20.5 | 3.8 | 5.7 |
|  |  | 2 | 12.2 | 3.0 | 4.1 |
| 2 | 10 | 0.5 | 10.3 | 4.8 | 6.2 |
|  |  | 1 | 7.3 | 3.8 | 5.3 |
|  |  | 2 | 5.4 | 3.0 | 9.2 |
| 2 | 30 | 0.5 | 8.4 | 4.8 | 10.2 |
|  |  | 1 | 6.4 | 3.8 | n/a |
|  |  | 2 | 4.9 | 3.0 | n/a |

Table 1: Diffusion, Gradient strength, and $T_2^*$ were varied to create a lookup table using equations 10 and 15. It is clear the resolution broadening due to frequency encoding is much more than phase encoding with the primary contribution being increased diffusion values. The $\Delta x_{nominal}$ for which R = 0.5 is also given in the final column to demonstrate the $\Delta x_{nominal}$ for which PE is twice as time efficient as the FE sequence. Note that n/a indicates there is no $\Delta x_{nominal}$ for which R = 0.5, meaning PE is more than 2x time efficient with the sample and hardware properties.

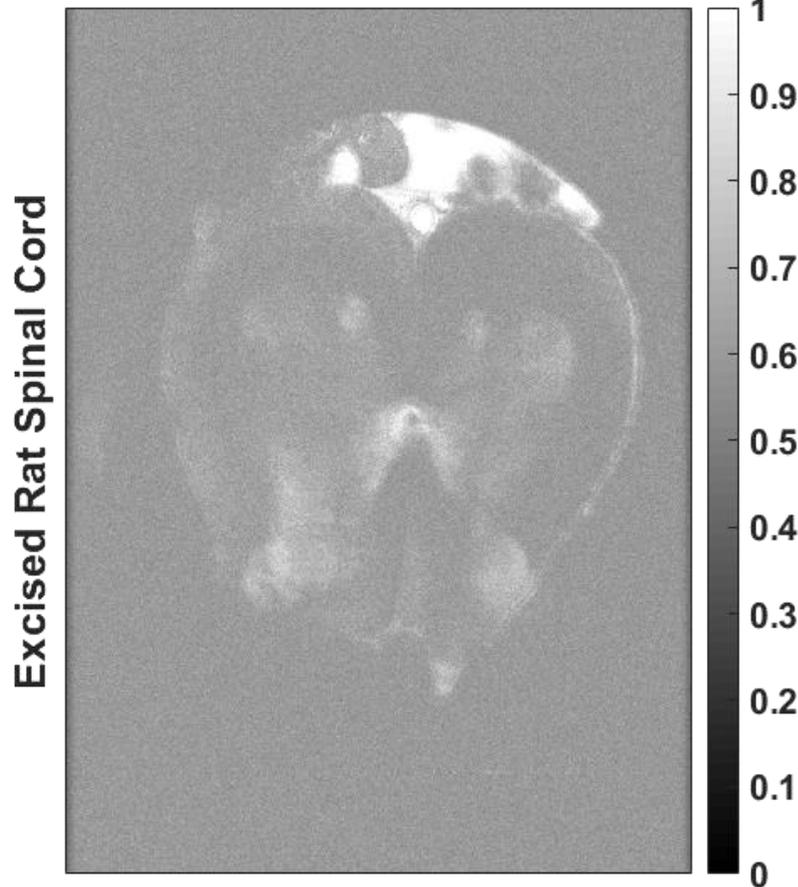 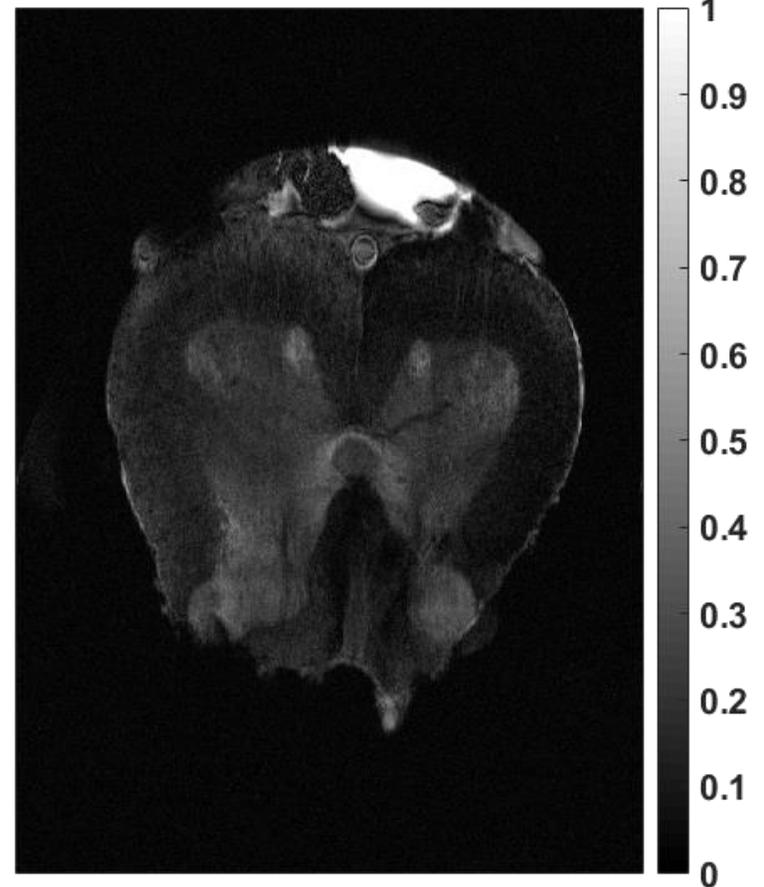

Figure 9: Upper sacral region of an excised rat spinal cord. In roughly 2 hours, PE provides a higher quality image with higher SNR and actual resolution. There is clear distinction between the white and gray matter. Vasculature can also be seen within the gray matter to the right and left of the central canal.

# 5. Discussion

The objective of this work is to give an MR microscopist tools to choose between PE and a conventional readout gradient. We lay out the hardware and sample properties in which the image quality and time efficiency of PE greatly exceeds that of a conventional readout sequence. Without diffusion present, the $\eta^{FE}$ is generally equal to $\eta^{PE}$. However, as voxels shrink and the diffusion of the sample attenuates signal during the frequency encoding readout gradient, PE offers benefits in the form of actual resolution and time efficiency especially in voxels < 15 µm and gradients < 2 T/m. This is important for future MRM experiments that are limited by finite gradient strengths and require time efficient acquisitions with resolution beyond 15 µm.

No work to date has included the SNR attenuation of diffusion in a side-by-side comparison of a readout and PE acquisition for MR Microscopy. Although Gravina and Cory discuss diffusion's effects in a SE, they do not include diffusion attenuation in their signal model and are primarily interested in imaging solids [14]. Choi and coworkers point out that constant time imaging may be more time efficient than SE in certain cases, however they do not generalize for PE or include variation of diffusion, $T_2^*$, and $G_{max}$ values as is included here.

The assumption that $t_{acq} = T_{read}$ in figure 2 is reasonable for certain imaging conditions. However, PE has highest SNR for $t_{acq} = 1.25 T_2^*$ which is not necessarily a realistic $T_{read}$ value given the $T_2^*$, limited gradient, and number of samples. A more realistic comparison was chosen for table 1, allowing for $T_{read}$ to be determined by the maximum gradient strength available. Figure 3 demonstrates the immediate advantage of PE in higher achievable resolution. However, the phase encoding resolution is dependent on the encoding time (or phase difference), broadening of resolution continues as $t_{enc}$ is lengthened for higher resolutions [11]. For frequency encoding the resolution depends on frequency difference between voxels so broadening stops at the diffusive limit (≈ 6 µm in figure 3) [12]. Therefore, at $\Delta x_{nominal}$ < 2 µm, phase encoding's $\Delta x_{actual}$ exceeds that of frequency encoding. This may be avoided with higher gradient strengths.

The method described in figure 4 provides the spatially dependent point spread function across the sample. With varying diffusivity and $T_2$ across the sample's resolution map, the FWHM is thus a $T_2$ and diffusion weighted contrast image. The large time commitment should be noted.

The PSF contrast image or resolution map may be of use for certain applications that are interested in multi-parametric variation with resolution.

Diffusion in the readout dimension has been derived and discussed in many previous publications [16,23,29,47,48]. The results in figures 4, 5, and 6 beg the question whether ADC values may be measured from a readout gradient by itself for instance by varying the $G_{read}/G_{dephase}$ ratio to vary the effective b-value. Although parsing out the attenuation's contributions from $T_2^*$ may be cumbersome, the advantage of this would be much shorter TR's, higher SNR, and more efficient diffusion maps since FLASH is inherently more efficient than SE [35].

In figure 6, the simulated readout gradient strengths may be used to calculate the SNR diffusion optimal bandwidth for a given resolution. Callaghan discusses this [21,24] however the methods presented here allow for a quick and relatively simple method of calculating the SNR optimal bandwidth considering diffusion. Given the target $\Delta x_{nominal}$, FOV, and sample properties, equation 16 may be used to find an optimal $T_{read}G_{read}$ combination or rather the diffusion optimal BW. However, a decrease in gradient strength for SNR may compromise the actual resolution of the image if the diffusive broadening exceeds the unavoidable contributions of sampling and relaxation. In figure 7, we demonstrate experimental verification that the imaging efficiency differences of FE and PE begin to become apparent at resolutions 10 – 15 μm especially with gradients < 2 T/m. Often pure phase encoding is quoted to be less SNR time efficient than readout acquisitions [13,14] which may lead some away from PE acquisitions. This may be true for single point imaging methods like constant time imaging, however when the sequence allows for longer acquisition times (i.e. $T_2^* > 1$ ms), PE is very comparable if not better than readout acquisitions especially at high resolution. Another added benefit not thoroughly expressed here is that since PE does not require a dephasing gradient, TE and $T_R$ may be shortened. Especially with longer $T_2^*$ values, most of the sequence duration could be spent acquiring signal, since phase encoding is the only time requirement between excitation and acquisition.

In figure 8, the measured PSF demonstrates clear agreement with phase encoding's resolution advantage. At 15 μm, the numerical PSF and measured PSF agree since most of the broadening is determined by the sampling function. However, as the resolution increases, the measured and analytical model differ. This is to be expected as the model doesn't include the extra phase encoding gradients that are used to measure the PSF as the diffusive broadening from these becomes more significant at higher resolution. The measured PSF however confirms that at high resolution, phase encoding provides finer resolution.

In figure 9, the excised rat spinal cord image is the highest reported in-plane resolution and smallest voxel size to date with the closest image being 23 x 23 x 300 µm resolution with a gradient echo in 3.5 hours [49].

The importance of this work may be demonstrated in equations 8-10 as summarized in table 1. There are clear transition points when PE is preferred in terms of image efficiency and actual resolution. For instance, $D = 2$ µm$^2$ ms$^{-1}$, $T_2^* = 10$ ms, $G_{max} = 0.5$ T/m resolution degradation for frequency encoding reaches 150% at $\Delta x_{nominal} = 10.3$ µm vs 4.8 µm for phase encoding. Essentially imaging features < 15 µm is impossible in the FE dimension without the use of phase encoding given these sample and hardware properties.

The benefit of PE is two-fold, SNR is relatively unaffected by diffusion while providing finer resolution than FE. Low $N_{acc}$ values have the added benefit of avoiding compounded systematic error which doesn't average out with increased accumulations [19].

# 6. Conclusion

In this work, we demonstrate the sequence, sample, and image properties which warrant switching from a FE to a PE acquisition. Diffusion during the readout gradient is experimentally demonstrated with shrinking voxel sizes. As has been previously shown in literature, the role of gradient strength in smaller voxels is demonstrated with simulated SNR and actual resolution. Conditions are laid out in which the imaging efficiency of PE greatly exceeds that of FE, an important guidepost for future high resolution MRM experiments. Theoretical resolution broadening in the readout and phase encoded dimensions is experimentally verified via measurement of the PSF. Finally, excised rat spinal cord images at 10 µm in-plane resolution demonstrate a practical scenario where PE exceeds FE's imaging efficiency and image quality at very high resolution.

# Acknowledgements

This work was generously supported by the Chan-Zuckerberg Initiative (CZI) for Deep Tissue Imaging. The authors would like to acknowledge Gary Drake for his help in designing the microcoil stage.

Appendix

  A. For simplification, $G_{read}$ is referred to as $G_r$ and $G_{dephase}$ is referred to as $G_d$. For readout gradients with varying $G_r$ to $G_d$ ratios, the integral describing the attenuation due to diffusion may be written as

$$A = exp\left[-\gamma^2 D \int_0^t \left(\int_0^{t'} G(t'')dt''\right)^2 dt'\right] \quad \text{A1}$$

Which is the solution to the Bloch-Torrey equation [39]. The gradient waveforms (fig. A1) may be described as

$$\int_0^{t'} G(t'')dt'' = -G_d t \quad \text{for } t < \tau \quad \text{A2}$$

$$\int_0^{t'} G(t'')dt'' = -G_d \tau + G_r(t - \tau) \quad \text{for } t > \tau$$

Substituting into the attenuation term and integrating until the center of the echo,

$$A = exp\left[-\gamma^2 D \left[\int_0^\tau G_d^2 t^2 dt + \int_\tau^{\tau + T_r/2} (-G_d \tau + G_r(t - \tau))^2 dt\right]\right] \quad \text{A3}$$

Referring to a.1, $G_r T_{read}/2 = G_d \tau$, therefore $\tau = G_r T_{read}/2G_d$. We can define some simplifying parameters to relate the term in the exponent to the readout gradient and duration. Defining

$$\beta = \frac{G_r}{G_d} \text{ and } \alpha = \frac{\beta + 1}{2} \quad \text{A4}$$

The echo time may be written in terms of $T_{read}$

$$\tau + \frac{T_{read}}{2} = \frac{G_r T_{read}}{2G_d} + \frac{T_{read}}{2} = T_{read}\left(\frac{G_r}{2G_d} + \frac{1}{2}\right) = T_{read}\alpha \quad \text{A5}$$

Performing the integrals in A3, substituting in equations a4, a5, and $\tau = G_r T_{read}/2G_d$ results in the exponent term in equation a1,

$$-\gamma^2 D G_r^2 T_{read}^3 \left[\alpha/3\left(\alpha^2 - \frac{3\alpha}{2} + \frac{3}{4}\right) + \alpha\beta\left(-\frac{\alpha}{2} + \frac{\beta}{4} + \frac{1}{2}\right) - \beta/24\left(\beta^2 + 3\beta + 2\right)\right] = \quad \text{A6}$$
$$-\gamma^2 D G_r^2 T_{read}^3 \varepsilon.$$

Where,

$$\varepsilon = \frac{\alpha}{3}\left(\alpha^2 - \frac{3\alpha}{2} + \frac{3}{4}\right) + \alpha\beta\left(-\frac{\alpha}{2} + \frac{\beta}{4} + \frac{1}{2}\right) - \frac{\beta}{24}(\beta^2 + 3\beta + 2). \quad \text{A7}$$

At β = 1, ε = 1/12, and at β = 2, ε = 1/8.

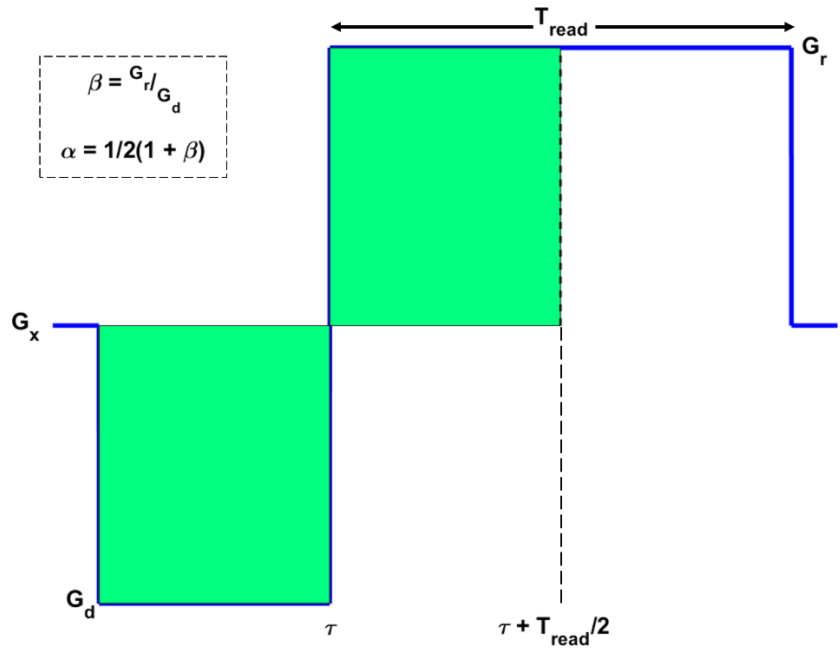

Figure A1: Gradient waveform for a gradient echo. The area under the dephase gradient is equal to ½ of the area under $G_r$ for conventional gradient echoes. TE is defined as the time between the center of the RF pulse and the center of $T_{read}$.

References

[1] P.C. Lauterbur, Image Formation by Induced Local Interactions: Examples Employing Nuclear Magnetic Resonance, Nat. 1973 2425394. 242 (1973) 190–191. https://doi.org/10.1038/242190a0.

[2] P. Mansfield, P.K. Grannell, NMR "diffraction" in solids?, J. Phys. C Solid State Phys. 6 (1973) L422. https://doi.org/10.1088/0022-3719/6/22/007.

[3] L.K. Hedges, Microscopic Nuclear Magnetic Resonance Imaging (NMR), State University of New York at Stony Brook, 1984.

[4] C.H. Lee, S.J. Blackband, P. Fernandez-Funez, Visualization of synaptic domains in the Drosophila brain by magnetic resonance microscopy at 10 micron isotropic resolution, Sci. Rep. 5 (2015) 8920. https://doi.org/10.1038/srep08920.

[5] C.H. Lee, N. Bengtsson, S.M. Chrzanowski, J.J. Flint, G.A. Walter, S.J. Blackband, Magnetic Resonance Microscopy (MRM) of Single Mammalian Myofibers and Myonuclei,

Sci. Reports 2017 71. 7 (2017) 1–9. https://doi.org/10.1038/srep39496.

[6]     J.J. Flint, C.H. Lee, B. Hansen, M. Fey, D. Schmidig, J.D. Bui, M.A. King, P. Vestergaard-Poulsen, S.J. Blackband, Magnetic resonance microscopy of mammalian neurons, Neuroimage. 46 (2009) 1037–1040. https://doi.org/10.1016/J.NEUROIMAGE.2009.03.009.

[7]     J.S. Morrow, Nuclear magnetic resonance imaging: a tool for microscopists?, J. Histochem. Cytochem. 34 (1986) 75–81. https://doi.org/10.1177/34.1.3941269.

[8]     J.B. Aguayo, S.J. Blackband, J. Schoeniger, M.A. Mattingly, M. Hintermann, Nuclear magnetic resonance imaging of a single cell, Nature. 322 (1986). https://doi.org/10.1038/322190a0.

[9]     H.Y. Chen, R. Tycko, Low-temperature magnetic resonance imaging with 2.8 μm isotropic resolution, J. Magn. Reson. 287 (2018) 47–55. https://doi.org/10.1016/J.JMR.2017.12.016.

[10]    L. Ciobanu, D.A. Seeber, C.H. Pennington, 3D MR microscopy with resolution 3.7 by 3.3 by 3.3 micron, J. Magn. Reson. 158 (2002) 178–182. https://doi.org/10.1016/S1090-7807(02)00071-X.

[11]    A.G. Webb, Optimizing the point spread function in phase-encoded magnetic resonance microscopy, Concepts Magn. Reson. Part A. 22A (2004) 25–36. https://doi.org/10.1002/CMR.A.20010.

[12]    E.W. McFarland, Time-independent point-spread function for NMR microscopy, Magn. Reson. Imaging. 10 (1992) 269–278. https://doi.org/10.1016/0730-725X(92)90486-J.

[13]    S. Choi, X.W. Tang, D.G. Cory, Constant time imaging approaches to NMR microscopy, Int. J. Imaging Syst. Technol. 8 (1997) 263–276. https://doi.org/10.1002/(SICI)1098-1098(1997)8:3<263::AID-IMA4>3.0.CO;2-8.

[14]    S. Gravina, D.G. Cory, Sensitivity and Resolution of Constant-Time Imaging, J. Magn. Reson. Ser. B. 104 (1994) 53–61. https://doi.org/10.1006/JMRB.1994.1052.

[15]    S. Emid, J.H.N. Creyghton, High resolution NMR imaging in solids, Phys. B+C. 128 (1985) 81–83. https://doi.org/10.1016/0378-4363(85)90087-7.

[16]    P. Glover, P. Mansfield, Limits to magnetic resonance microscopy, Reports Prog. Phys.

65 (2002) 1489. https://doi.org/10.1088/0034-4885/65/10/203.

[17] R. van Schadewijk, J.R. Krug, D. Shen, K.B.S. Sankar Gupta, F.J. Vergeldt, T. Bisseling, A.G. Webb, H. Van As, A.H. Velders, H.J.M. de Groot, A. Alia, Magnetic Resonance Microscopy at Cellular Resolution and Localised Spectroscopy of Medicago truncatula at 22.3 Tesla, Sci. Rep. 10 (2020). https://doi.org/10.1038/s41598-020-57861-7.

[18] D.I. Hoult, P.C. Lauterbur, The sensitivity of the zeugmatographic experiment involving human samples, J. Magn. Reson. 34 (1979) 425–433. https://doi.org/10.1016/0022-2364(79)90019-2.

[19] O. Nalcioglu, Z.H. Cho, Limits to signal-to-noise improvement by FID averaging in NMR imaging, Phys. Med. Biol. 29 (1984) 969–978. https://doi.org/10.1088/0031-9155/29/8/005.

[20] L. Ciobanu, A.G. Webb, C.H. Pennington, Magnetic resonance imaging of biological cells, Prog. Nucl. Magn. Reson. Spectrosc. 42 (2003) 69–93. https://doi.org/10.1016/S0079-6565(03)00004-9.

[21] Paul T. Callaghan, Principles of Nuclear Magnetic Resonance Microscopy, Clarendon Press, Oxford, 1991.

[22] K.M. Koch, D.L. Rothman, R.A. de Graaf, Optimization of static magnetic field homogeneity in the human and animal brain in vivo, Prog. Nucl. Magn. Reson. Spectrosc. 54 (2009) 69–96. https://doi.org/10.1016/j.pnmrs.2008.04.001.

[23] P.T. Callaghan, C.D. Eccles, Diffusion-limited resolution in nuclear magnetic resonance microscopy, J. Magn. Reson. 78 (1988) 1–8. https://doi.org/10.1016/0022-2364(88)90151-5.

[24] P.T. Callaghan, C.D. Eccles, Sensitivity and resolution in NMR imaging, J. Magn. Reson. 71 (1987) 426–445. https://doi.org/10.1016/0022-2364(87)90243-5.

[25] P.T. Callaghan, L.C. Forde, C.J. Rofe, Correlated Susceptibility and Diffusion Effects in NMR Microscopy Using both Phase-Frequency Encoding and Phase-Phase Encoding, J. Magn. Reson. Ser. B. 104 (1994) 34–52. https://doi.org/10.1006/JMRB.1994.1051.

[26] C.D. Eccles, P.T. Callaghan, High-resolution imaging. The NMR microscope, J. Magn. Reson. 68 (1986) 393–398. https://doi.org/10.1016/0022-2364(86)90261-1.

[27] P.T. Callaghan, Susceptibility-limited resolution in nuclear magnetic resonance microscopy, J. Magn. Reson. 87 (1990) 304–318. https://doi.org/10.1016/0022-2364(90)90007-V.

[28] Z.H. Cho, C.B. Ahn, S.C. Juh, H.K. Lee, R.E. Jacobs, S. Lee, J.H. Yi, J.M. Jo, Nuclear magnetic resonance microscopy with 4-μm resolution: Theoretical study and experimental results, Med. Phys. 15 (1988) 815–824. https://doi.org/10.1118/1.596287.

[29] C.B. Ahn, Z.H. Cho, A generalized formulation of diffusion effects in micron resolution nuclear magnetic resonance imaging, Med. Phys. 16 (1989) 22–28. https://doi.org/10.1118/1.596393.

[30] Z.H. Cho, C.B. Ahn, S.C. Juh, J.M. Jo, R.M. Friedenberg, R.E. Jacobs, S.E. Fraser, Recent progress in NMR microscopy towards cellular imaging, Philos. Trans. R. Soc. London. Ser. A Phys. Eng. Sci. 333 (1990) 469–475. https://doi.org/10.1098/rsta.1990.0174.

[31] P. Latta, M.L.H. Gruwel, V. Volotovskyy, M.H. Weber, B. Tomanek, Single-point imaging with a variable phase encoding interval, Magn. Reson. Imaging. 26 (2008) 109–116. https://doi.org/10.1016/J.MRI.2007.05.004.

[32] D. Xiao, B.J. Balcom, BLIPPED (BLIpped Pure Phase EncoDing) high resolution MRI with low amplitude gradients, J. Magn. Reson. 285 (2017) 61–67. https://doi.org/10.1016/J.JMR.2017.10.013.

[33] J.C. García-Naranjo, P.M. Glover, F. Marica, B.J. Balcom, Variable bandwidth filtering for magnetic resonance imaging with pure phase encoding, J. Magn. Reson. 202 (2010) 234–238. https://doi.org/10.1016/J.JMR.2009.11.006.

[34] E. Moore, R. Tycko, Micron-scale magnetic resonance imaging of both liquids and solids, J. Magn. Reson. 260 (2015) 1–9. https://doi.org/10.1016/J.JMR.2015.09.001.

[35] J. Link, J. Seelig, Comparison of deuterium NMR imaging methods and application to plants, J. Magn. Reson. 89 (1990) 310–330. https://doi.org/10.1016/0022-2364(90)90237-4.

[36] E.D. Becker, J.A. Ferretti, P.N. Gambhir, Selection of Optimum Parameters for Pulse Fourier Transform Nuclear Magnetic Resonance, Anal. Chem. 51 (1979) 1413–1420. https://doi.org/10.1021/AC50045A016/ASSET/AC50045A016.FP.PNG_V03.

[37]  T.L. James, A.R. Margulis, Biomedical magnetic resonance, (1984).

[38]  H.Y. Carr, E.M. Purcell, Effects of Diffusion on Free Precession in Nuclear Magnetic Resonance Experiments, Phys. Rev. 94 (1954) 630. https://doi.org/10.1103/PhysRev.94.630.

[39]  H.C. Torrey, Bloch Equations with Diffusion Terms, Phys. Rev. 104 (1956) 563. https://doi.org/10.1103/PhysRev.104.563.

[40]  M. Weiger, D. Schmidig, S. Denoth, C. Massin, F. Vincent, M. Schenkel, M. Fey, NMR microscopy with isotropic resolution of 3.0 µm using dedicated hardware and optimized methods, Concepts Magn. Reson. Part B Magn. Reson. Eng. 33B (2008) 84–93. https://doi.org/10.1002/CMR.B.20112.

[41]  R.W. Brown, Y.C.N. Cheng, E.M. Haacke, M.R. Thompson, R. Venkatesan, Magnetic Resonance Imaging: Physical Principles and Sequence Design: Second Edition, 2014. https://doi.org/10.1002/9781118633953.

[42]  M.D. Robson, J.C. Gore, R.T. Constable, Measurement of the point spread function in MRI using constant time imaging, Magn. Reson. Med. 38 (1997) 733–740. https://doi.org/10.1002/MRM.1910380509.

[43]  H. Zeng, R.T. Constable, Image distortion correction in EPI: comparison of field mapping with point spread function mapping, Magn. Reson. Med. 48 (2002) 137–146. https://doi.org/10.1002/MRM.10200.

[44]  R.M. Henkelman, Measurement of signal intensities in the presence of noise in MR images, Med. Phys. 12 (1985) 232–233. https://doi.org/10.1118/1.595711.

[45]  R.M. Henkelman, Erratum: Measurement of signal intensities in the presence of noise in MR images [Med. Phys. 12, 232 (1985)], Med. Phys. 13 (1986) 544–544. https://doi.org/10.1118/1.595860.

[46]  A. Macovski, Noise in MRI, Magn. Reson. Med. 36 (1996) 494–497. https://doi.org/10.1002/MRM.1910360327.

[47]  W.B. Hyslop, P.C. Lauterbur, Effects of restricted diffusion on microscopic NMR imaging, J. Magn. Reson. 94 (1991) 501–510. https://doi.org/10.1016/0022-2364(91)90136-H.

[48]  P.T. Callaghan, A. Coy, L.C. Forde, C.J. Rofe, Diffusive relaxation and edge

enhancement in NMR mxicroscopy, J. Magn. Reson. - Ser. A. 101 (1993) 347–350. https://doi.org/10.1006/JMRA.1993.1057.

[49] T. Weber, M. Vroemen, V. Behr, T. Neuberger, P. Jakob, A. Haase, G. Schuierer, U. Bogdahn, C. Faber, N. Weidner, In Vivo High-Resolution MR Imaging of Neuropathologic Changes in the Injured Rat Spinal Cord, AJNR Am. J. Neuroradiol. 27 (2006) 598. /pmc/articles/PMC7976991/ (accessed December 16, 2022).

[50] K.R. Minard, R.A. Wind, Solenoidal microcoil design. Part I: Optimizing RF homogeneity and coil dimensions, Concepts Magn. Reson. 13 (2001) 128–142. https://doi.org/10.1002/1099-0534(2001)13:2<128::AID-CMR1002>3.0.CO;2-8.

[51] K.R. Minard, R.A. Wind, Solenoidal microcoil design?Part II: Optimizing winding parameters for maximum signal-to-noise performance, Concepts Magn. Reson. 13 (2001) 190–210. https://doi.org/10.1002/cmr.1008.

Supplementary Material:

Supplementary Figure 1: Diffusion attenuates SNR with longer and stronger gradients. However, SNR also depends on the total time the receiver is on thus for some smaller voxel lengths, there are optimal dwell times for acquisitions Here Sample rate = 1/BW. See for example the 1-4 micron resolution in the plot on the left, where SNR peaks at certain dwell times. The plot on the right is the experimental measurement of this relationship. Notice how the change in SNR is not as drastic with longer dwell times as the voxel size is smaller. This is because diffusion is attenuating the signal.

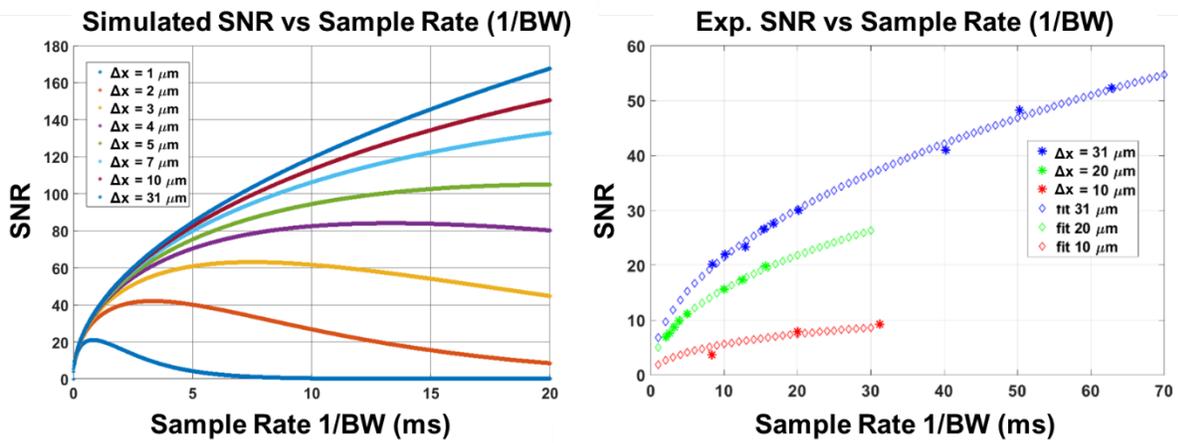

Supplementary figure 2: Micro-solenoid winding parameters including wire diameter (A) were optimized considering the transmit field homogeneity, sample size, and resistance of the circuit[50,51]. A stage for optimal placement of the sample and solenoid in the isocenter of the 15.2T magnet was designed (B and C). A circuit design was chosen to minimize resistance and common mode currents on the shield of the coaxial cable. A floating cable trap was also used to minimize any remaining common mode currents and ease the tune and match process.

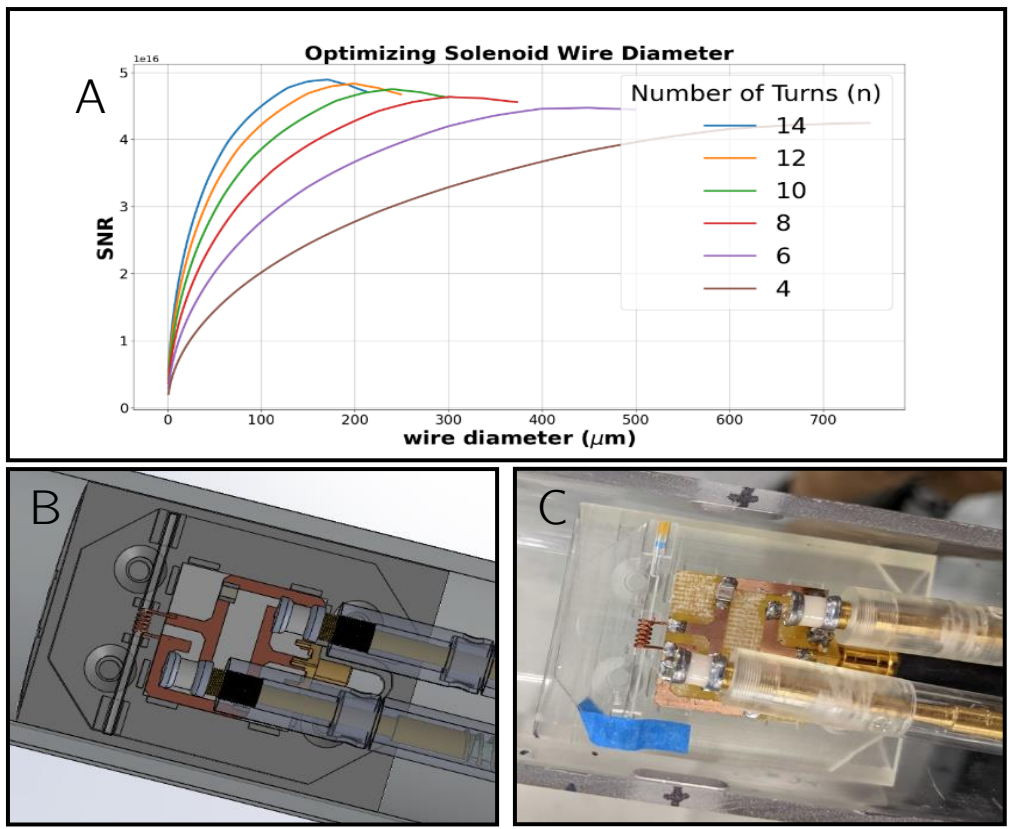

Supplementary figure 3: An Agar phantom was imaged at varying resolutions in the phase and frequency directions. The FWHM of the PSF was measured with a custom FLASH sequence. When the ratio of $N_{PSF}$ to $N_x$ is > 2, the FWHM shows good agreement with the theoretically predicted values given by McFarland. It should be noted that the $FOV_{PSF}$ = 0.75 FOV such that the sampling resolution in the PSF direction will always be 3/8 the resolution in the encoding direction when $N_{PSF}/N_x$ = 2. The measurement will be more accurate for higher PSF sampling resolutions however, $N_{PSF}/N_x$ = 2 is a compromise of FWHM measurement accuracy and the total time needed to acquire the data especially at very high resolution.

To determine an optimal $N_{PSF}/N_x$ ratio, a FLASH sequence's PSF with resolutions of 100 and 200 µm was measured with varying ratios $N_{PSF}/N_x$ from 1 to 6 (Supplementary figure 3). With $FOV_{PSF}$ = ¾ $FOV_x$ and a $N_{PSF}/N_x$ ratio of 2, a 200 µm resolution scan's measured resolution was ≈ McFarland's FWHM prediction including sampling, relaxation, and diffusion.

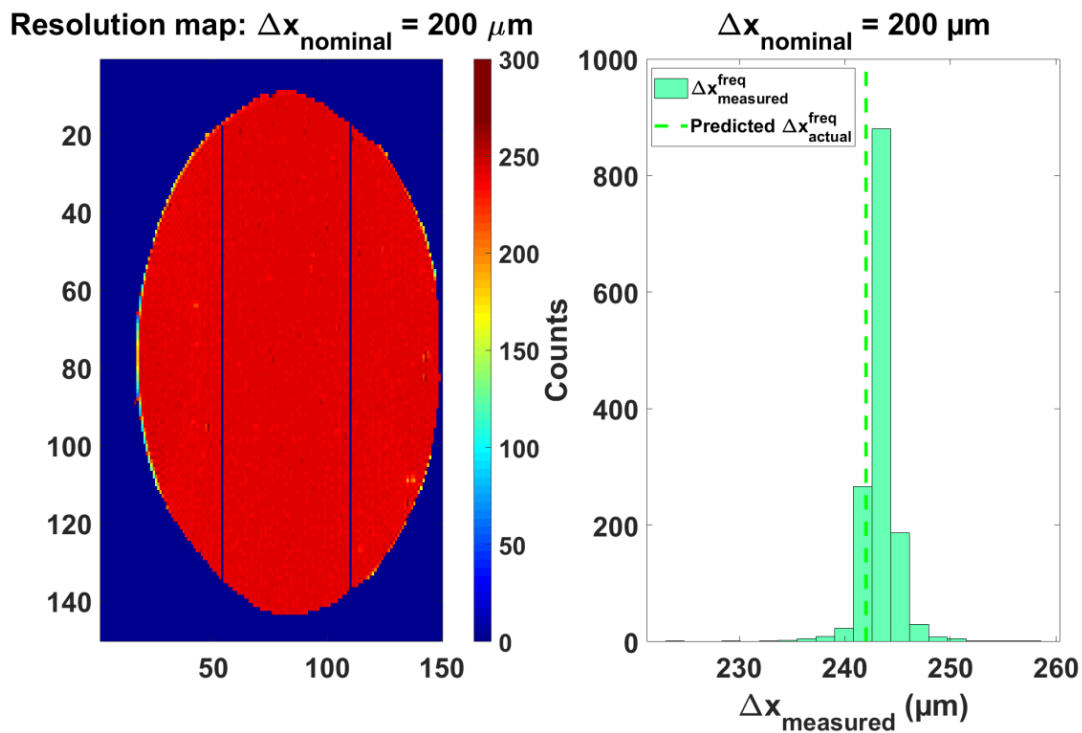

Supplementary figure 4: Equations 13 and 14 demonstrate the clear resolution benefit of phase encoding across many gradient strengths and diffusion values. Phase encoding may allow over 4 µm's worth of resolution considering diffusion alone. The actual resolution will be determined by including the sampling of the system, $T_2$ contributions, and the contributions of diffusion.

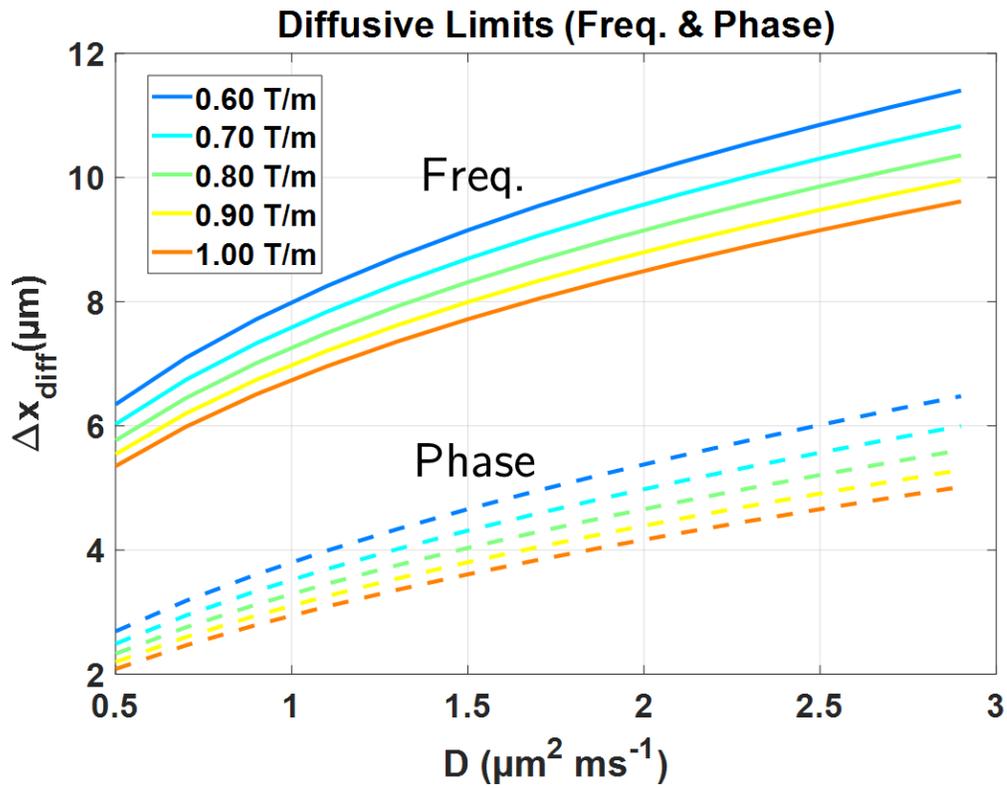